\documentclass[ reprint, amsmath, amssymb, aps, pra ]{revtex4-2}

\usepackage{graphicx}
\usepackage{bm}
\usepackage{amsmath,amssymb}
\usepackage{color}
\usepackage[caption=false]{subfig}
\usepackage{enumitem}

\usepackage{dcolumn}
\usepackage{hyperref}
\usepackage[mathlines]{lineno}

\usepackage{gensymb} 
\usepackage{makecell} 
\usepackage[normalem]{ulem}

\hypersetup{
    colorlinks,
    citecolor=black,
    filecolor=black,
    linkcolor=black,
    urlcolor=black
}

\begin{document}

\title{Fast, noise-free atomic optical memory with 35\% end-to-end efficiency}

\author{Omri Davidson}
\affiliation{Department of Physics of Complex Systems, Weizmann Institute of Science, Rehovot 7610001, Israel}
\author{Ohad Yogev}
\affiliation{Department of Physics of Complex Systems, Weizmann Institute of Science, Rehovot 7610001, Israel}
\author{Eilon Poem}
\affiliation{Department of Physics of Complex Systems, Weizmann Institute of Science, Rehovot 7610001, Israel}
\author{Ofer Firstenberg}
\affiliation{Department of Physics of Complex Systems, Weizmann Institute of Science, Rehovot 7610001, Israel}

\begin{abstract}
\noindent
Coherent optical memories will likely play an important role in future quantum communication networks. Among the different platforms, memories based on ladder-type orbital transitions in atomic gasses offer high bandwidth ($>100$ MHz), continuous (on-demand) readout, and low-noise operation. Here we report on an upgraded setup of our previously-reported fast ladder memory, with improved efficiency and lifetime, and reduced noise. The upgrade employs a stronger control field, wider signal beam, reduced atomic density, higher optical depth, annular optical-pumping beam, and weak dressing of an auxiliary orbital to counteract residual Doppler-broadening. For a 2 ns-long pulse, we demonstrate 53\% internal efficiency, 35\% end-to-end efficiency, $3\times 10^{-5}$ noise photons per pulse, and a $1/e$ lifetime of 108 ns. This combination of performances is a record for continuous-readout memories. 
\end{abstract}

\maketitle

\section{Introduction}
The generation of multi-photon states is highly desirable for quantum information processing applications such as photonic quantum computation and communication \cite{2007_Obrein_review_linear_optical_computing}. Quantum optical memories are likely to be critical components in such applications due to the probabilistic nature of photon sources and photonic gates \cite{2013_Nunn_enhancing_rates_with_memories}. 
Memories for classical and quantum light have been demonstrated in a plethora of systems, including cold atoms \cite{2005_Kuzmitch_cold_atoms,2015_Ding_cold_atoms}, hot atomic vapor \cite{2010_Nunn_vapor_memory,2022_Treutlein_vapor_memory}, solid-state artificial atoms \cite{2010_Sellars_cold_solid,2021_deRiedmatten_cold_solid}, and all-optical storage loops \cite{2017_Kwiat_storage_loop,2022_Silberhorn_storage_loop}. With the exception of storage loops, these memories can offer continuous (on-demand) readout, required for synchronizing continuously pumped photon sources \cite{photon_source_CW_hot_SebMoon_2016,Photon_source_paper,cavity_SPDC_Treutlein_2020}.

Memories based on the spin coherence of atomic ensembles, where both the signal photon and classical control field couple to the ground electronic orbital, have been extensively studied due to their potential for high storage efficiency \cite{2018_Chen_cold_atoms,2019_Du_cold_memory,2019_Zhang_hot_atoms} and long storage time \cite{2016_Buchler_cold_Atoms,2022_Namazi_hot_atoms,2018_Katz_vapor_memory}.
However, it is difficult to obtain noise-free operation in these memories due to four-wave mixing (FWM) noise \cite{2015_Nunn_warm_vapor_FWM_noise} and spontaneous Raman scattering from imperfect optical pumping \cite{2019_Zhang_hot_atoms}. These issues become even more pronounced, and eventually detrimental, when operating at high bandwidths that are comparable to the hyperfine splittings. Additionally, low noise operation requires filtering out the strong control field which typically requires etalon filters that reduce the signal transmission and hence the memory's end-to-end efficiency \cite{2018_Chen_cold_atoms,2019_Du_cold_memory,2019_Zhang_hot_atoms}.

An alternative to the spin memory is the orbital memory \cite{FLAME_paper,ORCA_ladder_memory,2022_Walmsley_ORCA_telecom}, where the excitation scheme is of a ladder-type, and the light is stored on a doubly-excited orbital. Although the memory lifetime is inherently limited by the lifetime of the orbital level, orbital memories are immune to FWM noise due to the ladder excitation scheme and the wavelength mismatch between the signal and control transitions  \cite{FLAME_paper,ORCA_ladder_memory}. Since the storage level is not populated even at high temperatures, the spontaneous Raman scattering noise is eliminated. Furthermore, the wavelength mismatch also enables the filtering of the strong control field from the retrieved signal field using conventional thin-film interference filters (IFs) with high transmission.

A fast ladder memory (FLAME) based on off-resonant cascaded absorption was demonstrated in cesium vapor by Kaczmarek \textit{et al.} \cite{ORCA_ladder_memory} and rubidium vapor by Thomas \textit{et al.} \cite{2022_Walmsley_ORCA_telecom}. In these demonstrations, the large wavelength mismatch between the signal and control fields results in substantial residual Doppler-broadening that limits the memory lifetime to a few ns. 
A much smaller mismatch exists for the ladder scheme $|5S_{1/2}\rangle \rightarrow |5P_{3/2}\rangle \rightarrow |5D_{5/2}\rangle$ in rubidium, for which the signal and control wavelengths are 780 nm and 776 nm, respectively. As demonstrated by Finkelstein \textit{et al.} \cite{FLAME_paper}, the reduced residual Doppler broadening of this cascaded transition enables a memory lifetime of $\sim 130$ ns in the absence of other decoherence mechanisms.

Here we report on an upgraded FLAME setup (FLAME-2) which has several improvements over the original demonstration (FLAME-1) \cite{FLAME_paper}. FLAME-2 uses an auxiliary, off-resonant dressing field that counteracts the residual Doppler broadening and further increases the memory lifetime \cite{Continuous_protection_paper}. To increase the memory efficiency and reduce the noise, FLAME-2 employs a longer vapor cell, which provides higher optical depth (OD) at a lower atomic density, and an annular optical pumping beam. Finally, the control field of FLAME-2 is stronger, which increases the memory efficiency for the on-resonance storage scheme, and the signal beam diameter is larger, reducing time-of-flight decoherence.
As part of the characterization of FLAME-2, we study on-resonance storage (also known as electromagnetically-induced-transparency storage \cite{2000_Fleischhauer_EIT_memory}) and off-resonance storage (pertaining to Raman storage \cite{2007_Nunn_Raman_storage}) of nanosecond-long pulses with variable width.


\begin{figure*} 
	\centering
	\includegraphics[width=\textwidth,trim=0.0cm 9.5cm 0.0cm 0.0cm ,clip=true] {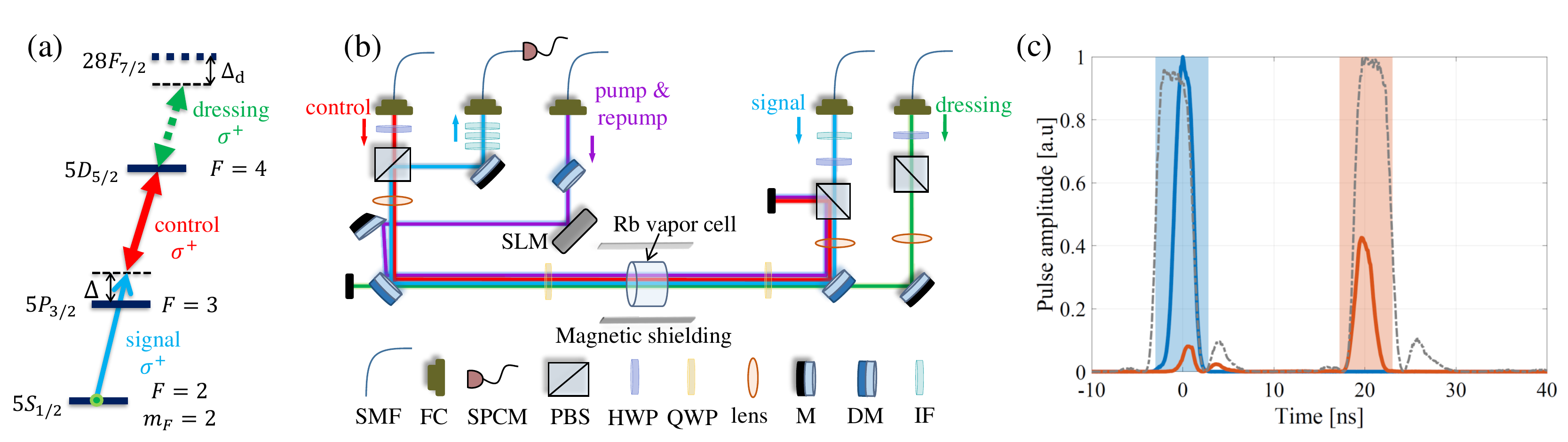}
	\caption{\textbf{FLAME-2 system.} 
	(\textbf{a}) The ladder level scheme of $^{87}\text{Rb}$. Initially, the atoms are optically pumped to the $|5S_{1/2},F=2,m_F=2\rangle $ ground state. The memory operates with all-orbital transitions to generate a coherence between the ground state and the doubly-excited state $|5D_{5/2},F=4,m_F=4\rangle $. The latter is off-resonantly coupled to the auxiliary state $|28F_{7/2}\rangle$ in order to compensate for the residual Doppler broadening of the $|5S_{1/2}\rangle \rightarrow |5P_{3/2}\rangle \rightarrow |5D_{5/2}\rangle$ transition.
	(\textbf{b}) The experimental setup. The signal and control beams counter-propagate inside the vapor cell. The pump and repump beams are reflected from an SLM to generate an annular beam inside the vapor cell and are inserted with a slight angle to the optical axis. The dressing beam is combined with the signal beam on a dichroic mirror.
	(\textbf{c}) Characteristic measurement with and without the memory operation. Shown are the reference signal pulse (blue), the stored and retrieved pulse for on-resonance storage (orange), and the control pulses (dashed-dotted gray). The shaded areas indicate the integration window size used for calculating the memory efficiency. 
	$\ $ DM: dichroic mirror, FC: fiber coupler, HWP: half-wave plate, IF: interference filter, M: mirror, PBS: polarizing beam splitter, QWP - quarter wave plate, SLM: spatial light modulator, SMF: polarization-maintaining single-mode fiber, SPCM: single-photon counting module.
	}
	\label{fig:experimental setup} 
\end{figure*}

\section{Experiment}

We use the $^{87}\text{Rb}$ level system shown in Fig.~\ref{fig:experimental setup}(a). 
The signal pulse with a wavelength of 780 nm couples the $|5S_{1/2},F=2\rangle \rightarrow |5P_{3/2},F=3\rangle$ transition with a detuning $\Delta = 0$ $(\Delta=1.1$ GHz) for on-resonance (off-resonance) storage. It originates from a continuous-wave (CW) distributed Bragg reflector (DBR) laser which is offset-locked to a stable master laser and passes through two amplitude electro-optics modulators (EOMs) to generate the signal pulses. We use an arbitrary waveform generator (PicoQuant PPG512) to generate Gaussian-like pulses with variable widths and stabilize the bias of the EOMs to achieve a pulse extinction ratio (ER) of $>1:10,000$. We attenuate the signal pulse intensity to an average of $\sim 0.1$ photons per pulse. 

The control field with a wavelength of 776 nm originates from a CW Ti:Sapphire laser, detuned from the $|5P_{3/2},F=3\rangle \rightarrow |5D_{5/2},F=4\rangle$ transition such that the two-photon transition $|5S_{1/2}\rangle \rightarrow |5P_{3/2}\rangle \rightarrow |5D_{5/2}\rangle$ is almost on-resonance. The storage and retrieval control pulses are generated with two free-space Pockels cells between cross-polarizers with a pulse ER of $>1:800$, and peak power at the vapor cell of 1.4 W. The $10\%-90\%$ rise and fall time of the pulses are $\sim 1.2$ ns, and the width of the pulses is variable. The timing of the signal and control pulses is controlled by a digital delay generator with a 10 ps timing resolution which operates the experimental sequence at a repetition rate of $10^5$ cycles per second.

The auxiliary dressing field originates from an external cavity diode laser at 1274 nm. It is initially amplified by an O-Band booster optical amplifier and further amplified by a tapered amplifier (TA). It is detuned by $\Delta_\text{d} = -570$ MHz from the $|5D_{5/2}\rangle \rightarrow |28F_{7/2}\rangle$ transition, and its power is 135 mW.
Lastly, the pump and repump fields, used to optically pump the atoms to the maximal spin state $|5S_{1/2},F=2,m_F=2\rangle$, have a wavelength of 795 nm and couple the $|5S_{1/2},F=2\rangle \rightarrow |5P_{1/2},F=2\rangle$ and $|5S_{1/2},F=1\rangle \rightarrow |5P_{1/2},F=2\rangle$ transitions, respectively [not shown in Fig.~\ref{fig:experimental setup}(a)]. They originate from DBR lasers amplified by TAs and have a power of 370 mW (105 mW) for pump (repump) before the vapor cell. The pump (repump) detuning is $-200$ MHz ($+150$ MHz). 
The frequencies of the control and dressing (pump and repump) lasers are stabilized using a wavelength meter with a resolution of $1$ MHz ($10$ MHz).

Figure~\ref{fig:experimental setup}(b) shows the experimental setup. All fields are $\sigma^+$ polarized inside the medium. The signal field is focused to a $1/e^2$ waist radius of $w_0 = 110 \ \mu$m inside a 25-mm-long isotopically purified $^{87}\text{Rb}$ vapor cell. The cell is placed inside three layers of mu-metal magnetic shielding and heated to $\sim 65 $ \degree C. The control field counter-propagates the signal with a waist radius of $w_0 = 180 \ \mu$m, and the dressing field co-propagates with the signal and has a waist radius of $w_0 = 210 \ \mu$m. The Rabi frequencies of the control and dressing fields (corresponding to 1.4 W and 135 mW) are $\Omega_\text{c}=640\pm 50$ MHz and $\Omega_\text{d}=30\pm 5$ MHz, respectively. Errors of system characterization represent 1 standard deviation (s.d.) of measurement uncertainty. 

The counter-propagating signal and control fields render the cascaded (two-photon) transition nearly Doppler-free. Nevertheless, to further suppress the residual Doppler broadening caused by their small wavelength difference, we employ a `continuous protection' scheme previously developed in our lab \cite{Continuous_protection_paper}. Here, the velocity-dependent Doppler shift is continuously compensated by velocity-dependent light shift, induced by the auxiliary dressing field. 

The pump and repump fields are reflected from a spatial light modulator (SLM) to generate an annular beam  with internal and external diameters of $400\ \mu$m and $1$ mm, respectively. This beam is inserted into the vapor cell with a D-shaped mirror at an angle of $\sim 0.85\degree$ to the optical axis, counter-propagating with respect to the signal field. The optical pumping efficiency of the $F=2$ ground level to the maximal spin state ($m_F=2$) is $94\pm2 \%$.

The output signal is sent to a single-photon counting module through a single-mode, polarization-maintaining fiber. There is no need to use etalon filters to clean the signal, due to the wavelength difference between the signal and control fields. Instead, we use two narrow-band IFs with 3-nm full-width at half-maximum (FWHM) in the output signal channel to filter out the residual control, pump, and repump fields, and a short-pass IF to filter out the residual dressing field. Additionally, we have one IF in the input signal channel, which prevents residual control light from coupling into the signal input fiber; This light might undergo spontaneous Raman scattering inside the fiber \cite{agrawal_2021_for_fiber_Raman_scattering}, which is frequency shifted and therefore partially transmitted through the IF of the output signal. Lastly, we use a dichroic mirror in the pump-repump input channel to block residual 780-nm light emitted from the TAs.
The transmission of the signal field through the memory setup is $66\pm 2 \%$, comprised of transmission through the vapor cell ($88\pm 1$\%), transmission through the optical filters ($94\pm 1$\%), coupling to the output fiber ($88\pm 2$\%), and the transmission of all other optical elements ($89\pm 2$\%). In the off-resonance storage, the transmission is further reduced by $15 \pm 2$\% due to absorption by residual $^{85}\text{Rb}$ inside the vapor cell.

Figure~\ref{fig:experimental setup}(c) shows a typical measurement of the memory experiment. Initially, we measure the signal pulse (count histogram shown in blue) when it is far-detuned from the atomic absorption line. Next, we turn on the control pulses (dashed-dotted gray) and again measure the signal output (orange; here shown for on-resonance storage). Throughout the paper, except for Fig.~\ref{fig:signal bandwidth}, the signal pulse is a Gaussian pulse with FWHM of 2 ns. The shaded areas in Fig.~\ref{fig:experimental setup}(c) (light blue and light orange) are the integration windows (6-ns long) used to calculate the total photon number in the reference and retrieved pulses from which we calculate the memory efficiency.  For the noise measurements, we block the signal input and integrate the photon counts in the same time window as for the retrieved pulse.

\section{Results}

\begin{figure} 
	\centering
	\includegraphics[width=\columnwidth,trim=0.0cm 0.0cm 0.0cm 0.5cm ,clip=true] {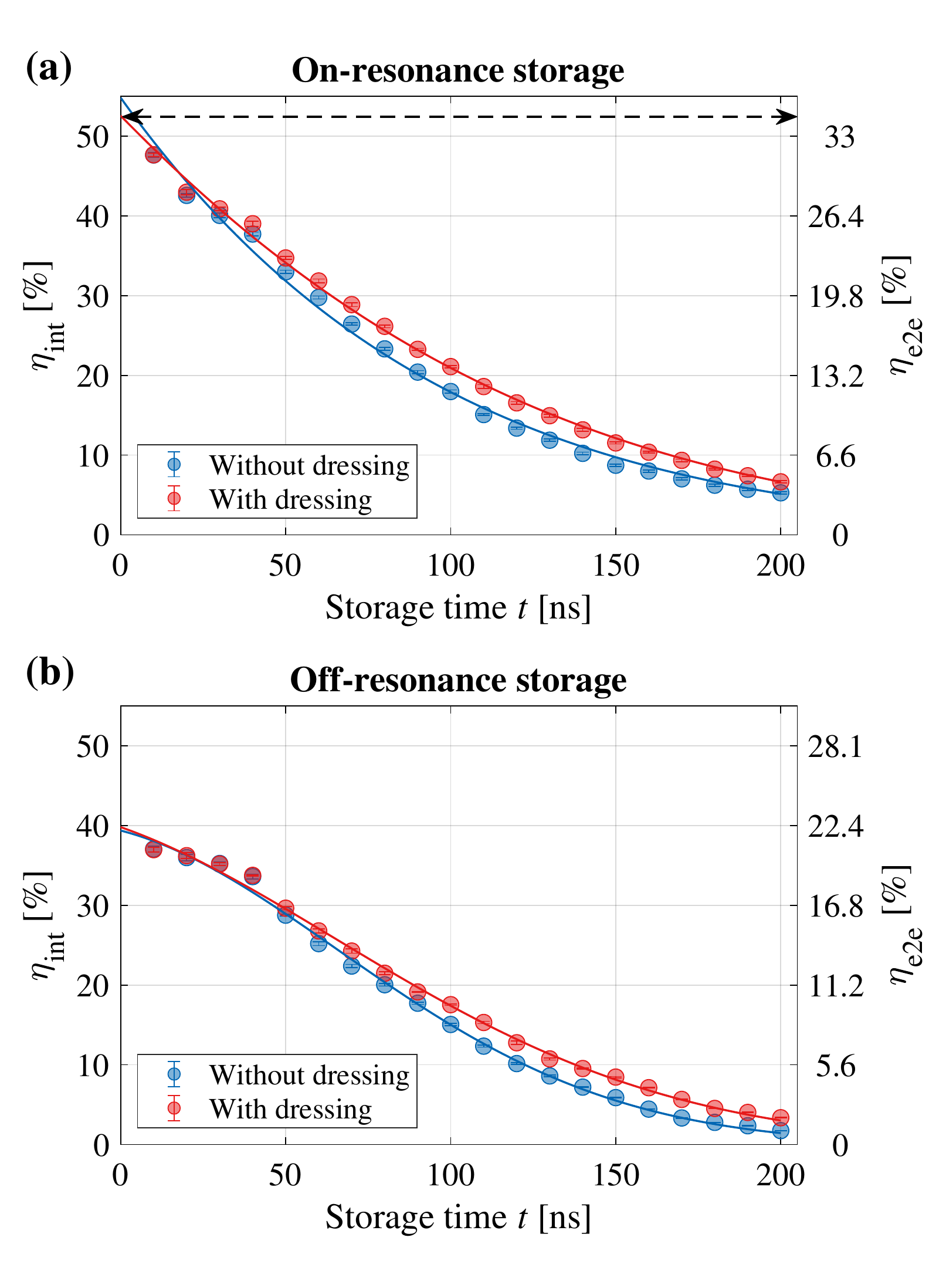}
	\caption{\textbf{Memory efficiency.} 
	(\textbf{a} and \textbf{b}) Efficiency (storage and retrieval) versus the storage time for (\textbf{a}) on-resonance and (\textbf{b}) off-resonance storage, without the dressing field (blue) and with the dressing field (red). 
	The left axis is the internal memory efficiency $(\eta_{\text{int}})$, and the right axis is the memory end-to-end efficiency $(\eta_{\text{e2e}})$ that includes the signal transmission through the entire memory setup. Circles are the measured data; the standard deviation of the mean of repeated measurements is smaller than the circles' size. The lines are fits to a model comprising exponential and Gaussian decays. The dashed arrow in (\textbf{a}) shows the zero-time efficiency.
	}
	\label{fig:Dressing} 
\end{figure}

We begin the memory characterization by optimizing the length of the control pulse, the two-photon detuning, and the OD to achieve maximal storage and retrieval efficiency. We find the optimal control FWHM to be 4 ns (3 ns) for on-resonance (off-resonance) storage. We note that further increasing the control pulse length reduces the memory efficiency, which might not be expected  for on-resonance storage. We attribute this to a reduction in the control pulse amplitude for longer pulses in our system.
The optimal two-photon detuning is found at slightly negative values of $-50$ MHz ($-20$ MHz), providing a relative efficiency increase of $2\pm1\%$ ($1\pm1$\%) for on-resonance (off-resonance) storage. In the off-resonance storage, this nonzero optimal detuning is due to control-induced light-shifts. In the on-resonance storage, it is due to the residual absorption of un-pumped atoms to the Doppler-broadened $|5P_{3/2},F=1,2\rangle $ states. 
We set the vapor temperature such that the OD, measured on the $|5S_{1/2},F=2\rangle \rightarrow |5P_{3/2},F=3\rangle$ transition, is $\text{OD}=19\pm 1$. Further increasing the OD saturates the storage efficiency while increasing the noise induced by the optical pumping beams.

Figure~\ref{fig:Dressing} shows the memory efficiency versus storage time $t$ with and without the dressing field for on-resonance and off-resonance storage. We find that the memory efficiency is higher for on-resonance storage when there is sufficient control power. This is to be expected, primarily as our control pulses are not chirped, as required for optimal off-resonance storage \cite{Gorshkov_free_space_model}.

\begin{figure} 
	\centering
	\includegraphics[width=\columnwidth,trim=0.0cm 0.0cm 0.0cm 0.0cm ,clip=true] {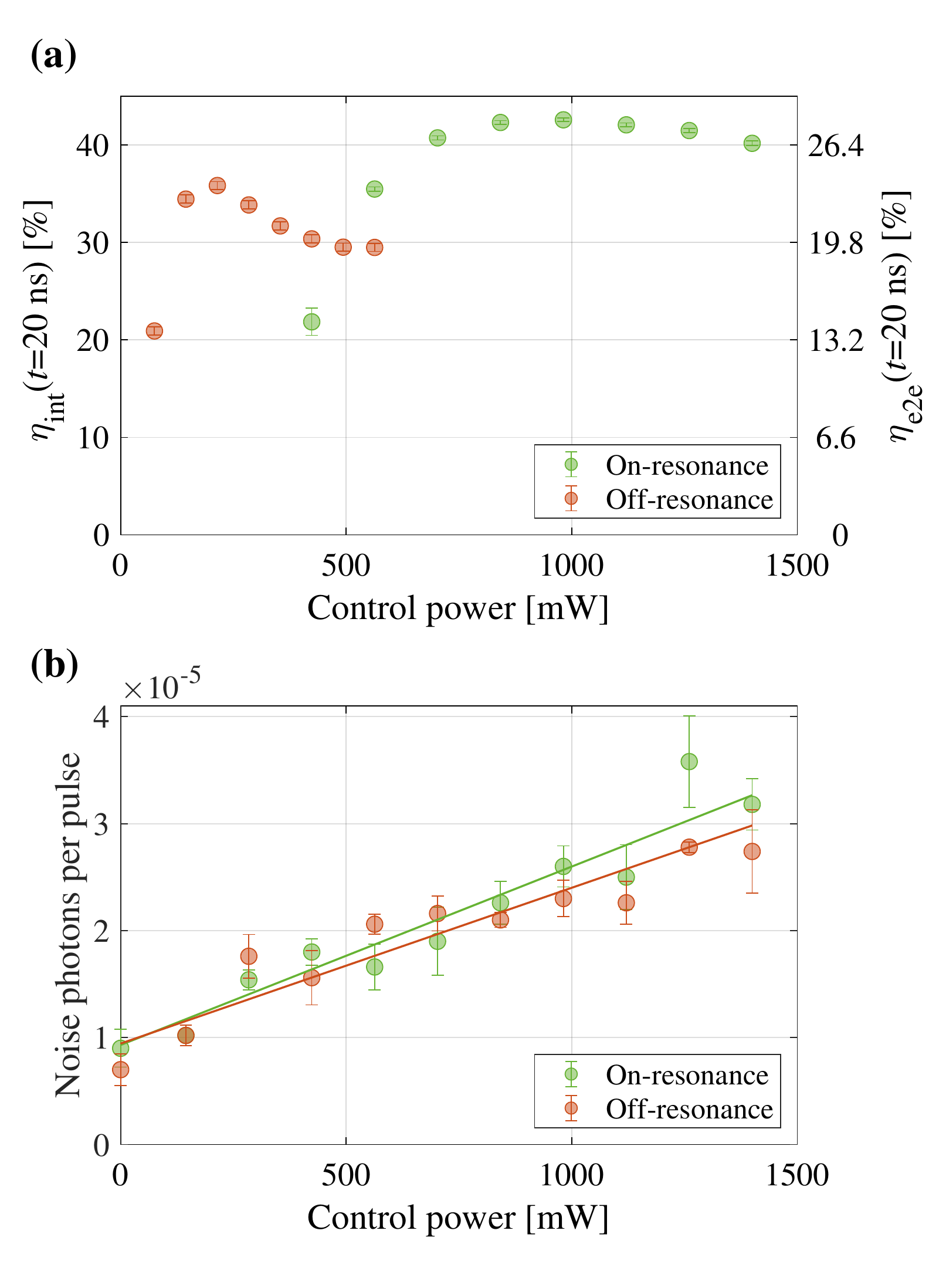}
	\caption{\textbf{Dependence of efficiency and noise on the control power.}
	(\textbf{a}) The memory efficiency at a storage time of 20 ns. Note that here and in Fig.~\ref{fig:signal bandwidth}, the $\eta_{\text{e2e}}$ axis refers to the end-to-end efficiency of the on-resonance storage; the off-resonance storage efficiency is 15\% lower due to absorption by residual $^{85}\text{Rb}$ atoms.   
	(\textbf{b}) The number of noise photons per retrieved pulse. The lines are a linear fit. 
	Results are shown for (green) on-resonance storage and (orange) off-resonance, both with the dressing field.
	}
	\label{fig:control power} 
\end{figure}

Adding the dressing field imparts a velocity-dependent light-shift which counter-acts the motional decoherence due to the residual Doppler broadening of the $|5S_{1/2}\rangle \rightarrow |5P_{3/2}\rangle \rightarrow |5D_{5/2}\rangle$ transition \cite{Continuous_protection_paper}. The dressing is kept constantly on, and we observe no increase in the homogeneous decoherence rate within the measurement uncertainty. The benefit of the dressing field is most significant at long storage times ($t>80$ ns), where it increases the memory efficiency by $>10\%$.

We model the memory efficiency versus time as $\eta(t) = \eta(0) e^{-t^2/2\tau_\sigma^2 - t/\tau_\gamma}$, with inhomogeneous ($\tau_\sigma$) and homogeneous ($\tau_\gamma$) decoherence times. We extract the memory $1/e$ lifetime $\tau_s$ from $\eta(\tau_s)=\eta(0)e^{-1}$ (see Ref.~\cite{FLAME_paper} for details). Using the dressing field increases the $1/e$ lifetime from $\tau_s = 90\pm 3$ ns ($102\pm2$ ns) to $\tau_s = 108\pm2$ ns ($113\pm 2$ ns) for on-resonance (off-resonance) storage, while keeping the zero-time efficiency $\eta(0)$ almost unchanged. With the dressing field, we measure an internal (end-to-end) short-time memory efficiency of $\eta_{\text{int}}(0)=52.6\pm 0.8 \%$ [$\eta_{\text{e2e}}(0)=34.7\pm 1.2 \%$] for the on-resonance storage, and $\eta_{\text{int}}(0)=39.8\pm 0.6 \%$ [$\eta_{\text{e2e}}(0)=22.3\pm 0.9 \%$] for the off-resonance storage. Here the errors are 1 s.d. of the fit uncertainty.

The dependence of the memory efficiency on the peak power of the control field is shown in Fig.~\ref{fig:control power}(a) for a storage time of 20 ns. For each control power, we optimize the control pulse timing and find that higher control powers require earlier timing to maximize the efficiency. On-resonance storage reaches higher efficiencies but necessitates a stronger control field, as it requires to generate a deep-enough transparency window within the Doppler-broadened absorption line. 

Figure~\ref{fig:control power}(b) shows the noise of the memory versus the control power. The noise per pulse caused by the pump beams is $\nu_\text{p} = 0.92\pm0.07 \times 10^{-5}$ photons. The noise originating from the control field increases linearly to a level of $\nu_\text{c}(\text{P}_\text{c}=1\text{ W})= 1.67 \pm 0.16 \times 10^{-5} $ photons for on-resonance storage and $\nu_\text{c}(\text{P}_\text{c}=1\text{ W}) = 1.46 \pm 0.14 \times 10^{-5} $ photons for off-resonance storage, where $\text{P}_\text{c}$ is the control power. This residual noise originates from reflection and perhaps a nonlinear frequency shift of the control field at the vapor cell facets.

\begin{figure} 
	\centering
	\includegraphics[width=\columnwidth,trim=0.0cm 0.0cm 0.0cm 0.0cm ,clip=true] {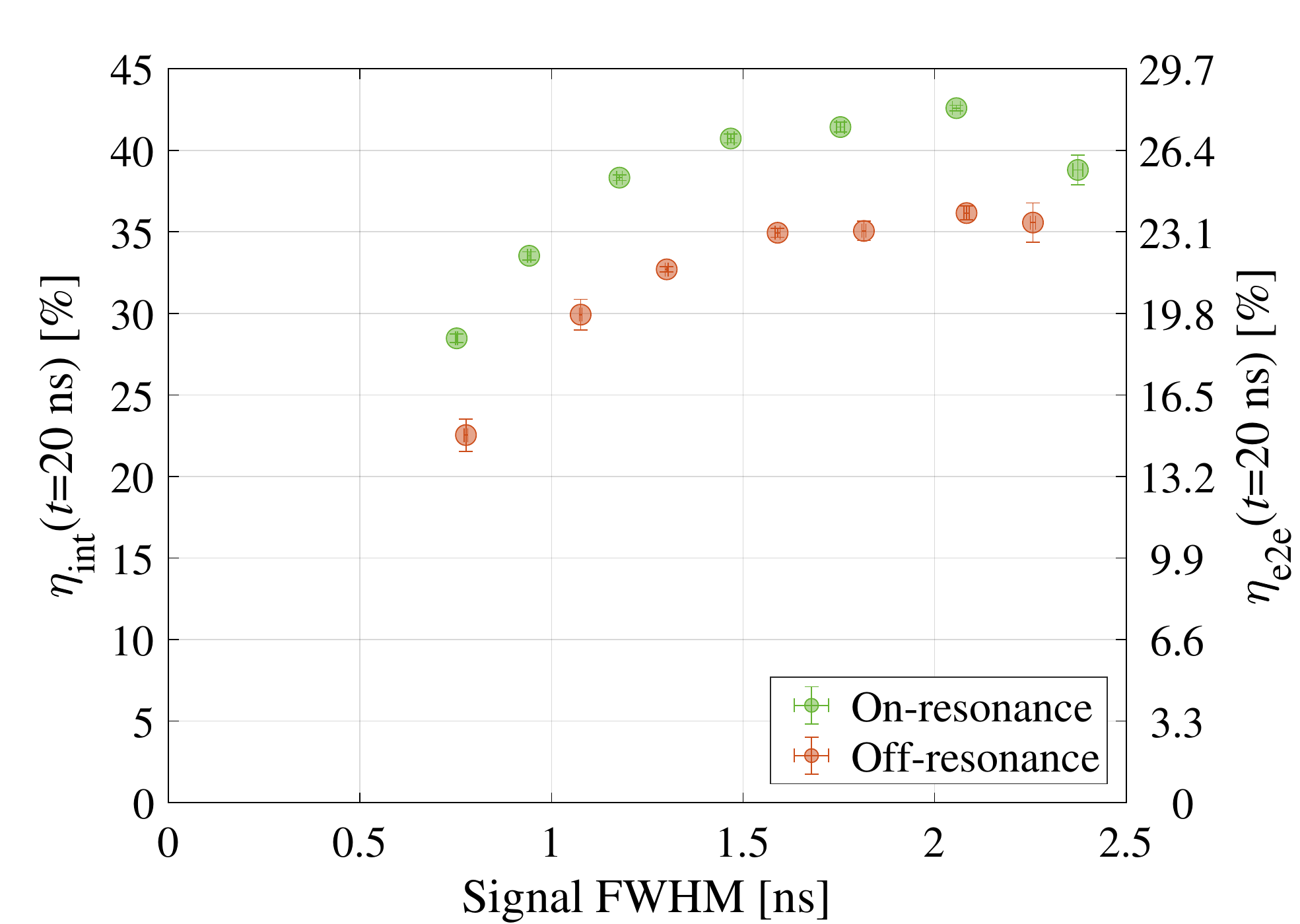}
	\caption{\textbf{Memory bandwidth.}
    The memory efficiency versus the signal pulse length is measured for a storage time of 20 ns with the dressing field. The efficiency is maximal for a 2-ns-long signal. 
	}
	\label{fig:signal bandwidth} 
\end{figure}

We study the memory bandwidth in Fig.~\ref{fig:signal bandwidth}. Here, for each signal pulse length, the control power and timing are optimized. For all measured signal pulses, the on-resonance storage has a higher efficiency than the off-resonance storage. As expected, the highest efficiency is obtained for the longer ($>1.5$ ns) signal pulses. 
While the efficiency reduces for shorter pulses, the effective fractional delay, which determines the potential enhancement of multi-photon rates in a synchronization application \cite{2013_Nunn_enhancing_rates_with_memories}, is actually increased when shortening the signal photons.

\section{Discussion}

\begin{table}[bt]
\caption{\textbf{Comparison of this work (FLAME-2) and Ref.~\cite{FLAME_paper} (FLAME-1).} We quote the performance of on-resonance storage for FLAME-2 and off-resonance storage for FLAME-1.}
\begin{tabular}{|l|c|c|c|c|}
\hline
   & \makecell{Internal \\ efficiency \\ $\eta_{\text{int}}(0) [\%]$}
   & \makecell{End-to-end \\ efficiency \\ $\eta_{\text{e2e}}(0) [\%]$ }
   & \makecell{ $1/e$ lifetime \\ $\tau_s$ [ns] }
   &  \makecell{ Noise \\ photons \\ $\nu \ [10^{-5}]$   }  \\ \hline
This work               & $52.6\pm 0.8$               & $34.7\pm 1.2$               & $108\pm2$     & $2.6\pm0.2$     \\ \hline
Ref.~\cite{FLAME_paper} & $32\pm1$                    & $25.1\pm0.8$                & $86\pm2$      & $5.9\pm0.4$     \\ \hline
\end{tabular}
\label{tab:FLAME comparison}
\end{table}

Table~\Ref{tab:FLAME comparison} compares the performance of the upgraded memory FLAME-2 to the original FLAME-1 \cite{FLAME_paper}. 
In FLAME-1, the off-resonance storage was slightly more efficient than the on-resonance storage, whereas in FLAME-2 the on-resonance storage is more efficient. This expected result is mainly due to the higher available control power. 
The efficiency is further improved in FLAME-2 by using a higher OD.
The noise in FLAME-2 is lower than in FLAME-1 due to the reduced atomic density in the longer vapor cell and the annular optical pumping beams.
The lifetime of FLAME-2 is increased by working with a larger signal beam that reduces the time-of-flight broadening and by employing the dressing field to counteract the residual Doppler broadening of the two-photon transition. 
Overall, FLAME-2 improves on FLAME-1 in all parameters. 

The end-to-end memory efficiency, storage time, and acceptance bandwidth are the key ingredients for enhancing photon synchronization rates. 
In our setup, the memory internal efficiency for on-resonance storage is limited firstly by the finite width of the control beam \cite{Nunn_theory_finite_control}, which is only $\sim 60\%$ wider than the signal beam, and by the imperfect optical pumping that causes residual absorption to the $|5P_{3/2},F=1,2\rangle$ states. By improving the optical pumping and increasing the control beam waist (while keeping the peak intensity the same), we estimate that an internal efficiency above $70\%$ is readily achievable for on-resonance storage. 
Further improving the memory internal efficiency will require replacing the Pockels cells that generate the control pulses in order to eliminate the after-pulse and optimize the control shape to the input signal pulse \cite{Gorshkov_optimal_storage_letter,Gorshkov_free_space_model}. This can be achieved by using an amplitude EOM that seeds a TA \cite{2022_Treutlein_vapor_memory}. 

The acceptance bandwidth of the memory is limited by the bandwidth of our control pulses. Using an amplitude EOM to generate the control pulses will thus enable, in principle, storage of shorter signal photons. However, this will also necessitate higher control powers, which are currently unavailable with tapered amplifiers.  
Finally, the end-to-end efficiency can be improved by increasing the setup transmission using a vapor cell with higher transmission and an output optical fiber with an anti-reflection coating. Realistically these can increase the overall transmission by about $10\%$.

In conclusion, we demonstrate an upgraded FLAME with high end-to-end efficiency, high bandwidth, and low noise. It outperforms the original FLAME demonstration in all of these parameters, and we outline a path for further improvements. The upgraded memory can be readily used to synchronize single photons with compatible wavelength and bandwidth, such as those generated in spontaneous cascaded emission \cite{photon_source_CW_hot_SebMoon_2016,Photon_source_paper}.

\begin{acknowledgments}
We thank Ran Finkelstein for helpful discussions.
We acknowledge financial support from the Israel Science Foundation, the US-Israel Binational Science Foundation (BSF) and US National Science Foundation (NSF), the Minerva Foundation with funding from the Federal German Ministry for Education and Research, the Estate of Louise Yasgour, and the Laboratory in Memory of Leon and Blacky Broder.
\end{acknowledgments}


\bibliographystyle{unsrt}

\bibliography{Memory_paper_bibliography.bib}

\end{document}